\documentclass[conference,a4paper]{APSIPA2026}
\usepackage{amsmath}
\usepackage{amssymb}
\usepackage{amsfonts}
\usepackage{graphicx}
\usepackage{multirow}
\usepackage{threeparttable}
\usepackage[backend=biber,style=ieee,]{biblatex}
\usepackage{url}
\usepackage{mathtools}
\usepackage{subcaption}
\usepackage{cleveref}

\newcommand{\ip}[2]{{#1}^{\top} {#2}}
\DeclareMathOperator*{\argmax}{arg~max}
\crefname{equation}{eq.}{eqs.}
\Crefname{equation}{Eq.}{Eqs.}
\crefname{figure}{fig.}{figs.}
\Crefname{figure}{Fig.}{Figs.}
\crefname{table}{table}{tables}
\Crefname{table}{Table}{Tables}

\usepackage{geometry}
\usepackage{fancyhdr}

\fancypagestyle{firststyle}{
  \fancyhf{}
  \fancyhead[C]{2026 Asia Pacific Signal and Information Processing Association Annual Summit and Conference (APSIPA ASC)}
}

\newcommand{\addieeecopyrightnotice}{%
  \AddToShipoutPictureFG*{%
    \AtPageLowerLeft{%
      \raisebox{6mm}{%
        \makebox[\paperwidth][c]{%
          \parbox{0.94\paperwidth}{%
            \centering
            \fontsize{6}{7}\selectfont
            \textcopyright~2026 IEEE. Personal use of this material is permitted.
            Permission from IEEE must be obtained for all other uses, in any current
            or future media, including reprinting/republishing this material for
            advertising or promotional purposes, creating new collective works, for
            resale or redistribution to servers or lists, or reuse of any copyrighted
            component of this work in other works.%
          }%
        }%
      }%
    }%
  }%
}

\begin{document}

\addieeecopyrightnotice

\title{Alignment of Similarity-Transformed Images \\
Based on Fourier--Mellin Transform \\
Using Auxiliary Function Method}

\author{
\authorblockN{
Shinji Yamashita\authorrefmark{1},
Yuma Kinoshita\authorrefmark{1}, and
Hitoshi Kiya\authorrefmark{2}
}

\authorblockA{
\authorrefmark{1}
Tokai University, Japan \\
E-mail: \{5CEIM050, ykinoshita\}@tokai.ac.jp}

\authorblockA{
\authorrefmark{2}
Tokyo Metropolitan University, Japan\\
E-mail: kiya@tmu.ac.jp}
}

\maketitle

\begin{abstract}
This paper proposes an algorithm for estimating the similarity transformation, namely translation, scale, and rotation, 
between two images with subpixel accuracy. 
Image registration is a fundamental technique for aligning images acquired under different viewpoints and imaging conditions, 
and a representative approach based on maximizing discrete cross-correlation is the Fourier--Mellin registration. 
However, the Fourier--Mellin approach often fails to achieve sufficient alignment accuracy
when subpixel-level estimation is required. 
The proposed method integrates (i) scale-and-rotation estimation from the Fourier magnitude spectrum in a log-polar representation and (ii) maximization of phase-only correlation based on the auxiliary function method. 
This integration enables a two-stage estimation procedure: it first estimates scale and rotation without being affected by translation, and then estimates translation with subpixel precision in the spatial domain using the corrected image pair. 
A simulation experiment on image pairs subjected to random similarity transformations demonstrates
that the proposed method reduces estimation errors in scale, rotation, and translation
compared with Fourier--Mellin-based registration methods using discrete cross-correlation. 
\end{abstract}

\section{Introduction}
Image alignment (registration) is a technique for aligning images by estimating the geometric transformation between them and compensating for it so that corresponding pixels (or feature points) coincide in a common coordinate system.
This paper focuses on similarity transformations, a class of geometric transformations consisting of translation, rotation, and scaling.
Image alignment is indispensable in a wide range of applications such as medical image processing~\cite{haskins2020deep}, panorama generation~\cite{matthew2007panorama}, 3D reconstruction~\cite{noah20083d}, and high dynamic range (HDR) imaging~\cite{reinherd2005hdr,hasinoff2016burst}.
Recently, there has been increasing demand for fast and highly accurate alignment on resource-constrained devices (e.g., smartphones), which highlights the importance of computationally efficient algorithms.

Image alignment methods can be broadly categorized into homography-based and intensity-based approaches.
In homography-based methods, correspondences are extracted using feature descriptors such as SIFT~\cite{lowe2004distinctive}, SURF~\cite{bay2006surf}, and A-KAZE~\cite{alcantarilla2012kaze}, and a homography between two images is typically estimated after outlier rejection via the RANSAC algorithm~\cite{fischler1981random}.
Recently, deep-learning-based homography estimation has also been investigated~\cite{cao2022iterative}.
In contrast, intensity-based methods widely employ direct optimization frameworks that estimate transformation parameters by directly minimizing pixel-wise errors, as well as optical-flow-based formulations~\cite{lucas1981iterative,dosovitskiy2015flownet,ilg2017flownet,jung2023anyflow}.
However, both homography estimation and optical flow estimation methods can impose substantial computational burdens in resource-limited environments.

To address the computational cost issue, two-dimensional cross-correlation-based methods~\cite{knapp1976generalized,takita2003highaccuracy,hasinoff2016burst} have been widely used for real-time image alignment.
These methods estimate translational displacement by maximizing the two-dimensional cross-correlation function between two images.
Cross-correlation for discrete pixel shifts can be computed efficiently using the fast Fourier transform (FFT).
Furthermore, Kinoshita \textit{et al.} proposed a computationally efficient algorithm that estimates the peak position of the cross-correlation function with subpixel accuracy based on an optimization algorithm known as the auxiliary function method or the majorization--minimization (MM) method~\cite{kinoshita2023auxliliary_function}.
However, these conventional methods assume that the displacement between images is purely translational and thus estimate only translation parameters.
In practical imaging scenarios, zooming and camera rotation often cause misalignment that includes scaling and rotation, for which a simple translation model is insufficient.

Therefore, there is a need for a highly accurate alignment algorithm that can handle similarity transformations while keeping computational cost low.
As a representative method that incorporates scaling and rotation, a Fourier--Mellin-transform-based registration is known~\cite{reddy1996fftbased}.
It exploits two properties. 
First, the magnitude spectrum is invariant to translation. 
Second, scaling and rotation in the spatial domain correspond to translations in the log-polar domain. 
Using these properties, it estimates translation, rotation, and scale through two rounds of discrete cross-correlation maximization.
However, because it also involves resampling due to the log-polar transform, maximizing discrete cross-correlation suffers from limited alignment accuracy.

To overcome this issue, this paper proposes a computationally efficient method for estimating all similarity transformation parameters with subpixel accuracy.
Specifically, we apply the high-precision correlation maximization algorithm based on the auxiliary function method~\cite{kinoshita2023auxliliary_function} to both (i) scale-and-rotation estimation based on correlation between magnitude spectra in the log-polar domain and (ii) translation estimation based on correlation between images in the spatial domain within the Fourier--Mellin framework.
This enables high-precision estimation of scale and rotation, followed by subpixel-accurate translation estimation on the corrected image pair.

A simulation experiment using five image pairs generated from each of three original images by applying random similarity transformations showed that the proposed method consistently reduced the absolute errors in scale, rotation, and translation compared with a Fourier--Mellin-based registration method using discrete cross-correlation.

\section{Preliminaries}
This section first formulates the alignment problem for similarity-transformed images.
It then reviews maximization of the two-dimensional cross-correlation function based on the auxiliary function method, which serves as a building block of the proposed approach.

\subsection{Problem formulation}
Let $x[\boldsymbol{p}]$ and $y[\boldsymbol{p}]$ denote the pixel values of two-dimensional discrete signals $x$ and $y$ at pixel $\boldsymbol{p}=(p_1,p_2)^\top$, respectively, where $\cdot^\top$ denotes transpose.
We consider the case where the misalignment between $x$ and $y$ is represented by a combination of translation, rotation, and scaling, i.e., $y$ is a similarity-transformed version of $x$.
Their relationship is given by
\begin{equation}
  y[\boldsymbol{p}]=x[sR(\theta)\boldsymbol{p}+\boldsymbol{\Delta p}] .
\end{equation}
Here, $\Delta\boldsymbol{p}\in\mathbb{R}^2$ denotes translation, $R(\theta)\in\mathbb{R}^{2\times2}$ is the rotation matrix
\begin{equation}
R(\theta)=
\begin{pmatrix}
\cos\theta & -\sin\theta\\
\sin\theta & \cos\theta
\end{pmatrix},
\end{equation}
and $s\in\mathbb{R}^+$ denotes the scale factor.
The goal is to estimate the unknown similarity transformation parameters $s$, $\theta$, and $\Delta\boldsymbol{p}$ from an image pair $(x,y)$, and to obtain the compensated image
$y''[\boldsymbol{p}] = y\!\left[\frac{1}{s}R^{-1}(\theta)(\boldsymbol{p}-\Delta\boldsymbol{p})\right]$.

\subsection{Maximization of two-dimensional cross-correlation}
\label{sec:aux}
In~\cite{kinoshita2023auxliliary_function}, Kinoshita \textit{et al.} proposed an algorithm for estimating the translation parameter $\Delta\boldsymbol{p}$ by maximizing a generalized two-dimensional cross-correlation function, focusing on the case where the displacement is purely translational (i.e., $s=1$ and $\theta=0$).
Their method first introduces a generalized two-dimensional cross-correlation function defined in the frequency domain as the objective function and then iteratively solves the resulting continuous maximization problem using the auxiliary function method.

Let $\hat{x}$ and $\hat{y}$ be the $N\times M$ two-dimensional discrete Fourier transforms (DFTs) of $x$ and $y$, respectively, and let the angular frequency vector be
$\boldsymbol{\omega}_{kl}=(\omega_k,\omega_l)^\top=(2\pi k/N,\ 2\pi l/M)^\top$.
Define the two-dimensional cross spectrum as
$\hat{\Phi}_2^{(xy)}(\boldsymbol{\omega}_{kl})=\hat{x}^*(\boldsymbol{\omega}_{kl})\hat{y}(\boldsymbol{\omega}_{kl})$.
Then, the generalized two-dimensional cross-correlation function is given by
\begin{equation}
  \check{\Phi}_2^{(xy)}(\boldsymbol{p})
  = \frac{1}{NM}\sum_{k\in K}\sum_{l\in L}
    w_{kl}\hat{\Phi}_2^{(xy)}(\boldsymbol{\omega}_{kl})
    \exp\!\left(j\ip{\boldsymbol{\omega}_{kl}}{\boldsymbol{p}}\right)
  \label{eq:gcc2_prep}
\end{equation}
where the summation ranges are
$K=\{-N/2+1,-N/2+2,\dots,N/2\}$ and $L=\{-M/2+1,-M/2+2,\dots,M/2\}$.
The weights $w_{kl}\in\mathbb{R}^+$ yield ordinary cross-correlation when $w_{kl}=1$, and phase-only correlation when $w_{kl}=|\hat{\Phi}_2^{(xy)}(\boldsymbol{\omega}_{kl})|^{-1}$~\cite{kuglin1975phase,knapp1976generalized}.
Allowing $\boldsymbol{p}$ in \Cref{eq:gcc2_prep} to vary continuously over $\mathbb{R}^2$ is equivalent to interpolating the discrete cross-correlation function using a two-dimensional periodic sinc function.
Using~\Cref{eq:gcc2_prep}, translation estimation is achieved by solving
\begin{equation}
  \tilde{\Delta\boldsymbol{p}}
  = \argmax_{\boldsymbol{p}\in\mathbb{R}^2}\check{\Phi}_2^{(xy)}(\boldsymbol{p}) .
  \label{eq:problem_translation}
\end{equation}

When $x$ and $y$ are strictly band-limited, by exploiting the conjugate symmetry of the cross spectrum,~\Cref{eq:gcc2_prep} can be rewritten as a linear combination of cosine functions with coefficients $\alpha_{kl}$:
\begin{equation}
  \check{\Phi}_2^{(xy)}(\boldsymbol{p})
  = \frac{1}{NM}\sum_{k\in K^+}\sum_{l\in L}
    \alpha_{kl}\cos\!\left(\ip{\boldsymbol{\omega}_{kl}}{\boldsymbol{p}}+\varphi_{kl}\right),
  \label{eq:objective_prep}
\end{equation}
where $K^+=\{0,1,\dots,N/2\}$ and $\varphi_{kl}=\angle \hat{\Phi}_2^{(xy)}(\boldsymbol{\omega}_{kl})$.
A quadratic function that lower-bounds this cosine sum is known to exist~\cite{yamaoka2019subsamplea}:
\begin{align}
  &\check{\Phi}_2^{(xy)}(\boldsymbol{p}) \ge
  Q(\boldsymbol{p},\boldsymbol{\theta}) \nonumber \\
  &= \sum_{k\in K^+}\sum_{l\in L}
    -\frac{\alpha_{kl}}{2NM}\frac{\sin\theta_{kl}}{\theta_{kl}}
    \left(\ip{\boldsymbol{\omega}_{kl}}{\boldsymbol{p}}+\varphi_{kl}+2n_{kl}\pi\right)^2 + C ,
  \label{eq:aux_prep}
\end{align}
where $\boldsymbol{\theta}=\{\theta_{kl}\}_{k,l}$ is a set of auxiliary variables, $C$ is a constant independent of $\boldsymbol{p}$, and $n_{kl}\in\mathbb{Z}$ satisfies
$|\ip{\boldsymbol{\omega}_{kl}}{\boldsymbol{p}}+\varphi_{kl}+2n_{kl}\pi|\le\pi$.

The method in~\cite{kinoshita2023auxliliary_function} iterates the following procedure to obtain a solution of~\Cref{eq:gcc2_prep}:
starting from an initial point $\boldsymbol{p}^{(i)}$, update $\boldsymbol{\theta}$ so that
$\check{\Phi}_2^{(xy)}(\boldsymbol{p}^{(i)}) = Q(\boldsymbol{p}^{(i)},\boldsymbol{\theta})$ holds, and then update $\boldsymbol{p}$ to a maximizer of $Q(\boldsymbol{p},\boldsymbol{\theta}^{(i+1)})$ using the updated $\boldsymbol{\theta}^{(i+1)}$.

The next section combines this framework with Fourier--Mellin-transform-based registration~\cite{reddy1996fftbased} to estimate similarity transformation parameters $s$, $\theta$, and $\Delta\boldsymbol{p}$, including rotation and scaling.

\section{Proposed method}
This paper proposes a subpixel-accurate method for estimating similarity transformation parameters within the Fourier--Mellin framework.
The key idea is to apply the auxiliary-function-based cross-correlation maximization method to both the scale-and-rotation estimation stage and the translation estimation stage.
Unlike the conventional Fourier--Mellin-based registration method, the proposed method replaces discrete cross-correlation maximization with continuous maximization based on the auxiliary function method.

The proposed method is based on two well-known properties of the Fourier magnitude spectrum.
First, the magnitude spectrum of an image is invariant to translation in the spatial domain.
Second, when the magnitude spectrum is represented in log-polar coordinates $(\rho,\phi)=(\log\|\boldsymbol{\omega}\|,\angle\boldsymbol{\omega})$, spatial-domain scaling and rotation correspond to translations along the $\rho$-axis and the $\phi$-axis, respectively.

Based on these properties, the conventional Fourier--Mellin-based approach first estimates the scale factor $s$ and rotation angle $\theta$ in the log-polar domain~\cite{reddy1996fftbased}.
It then estimates the translation $\Delta\boldsymbol{p}$ in the spatial domain after compensating for scale and rotation.

\begin{figure}[t]
  \begin{subfigure}{\linewidth}
      \centering
      \includegraphics[width=\linewidth]{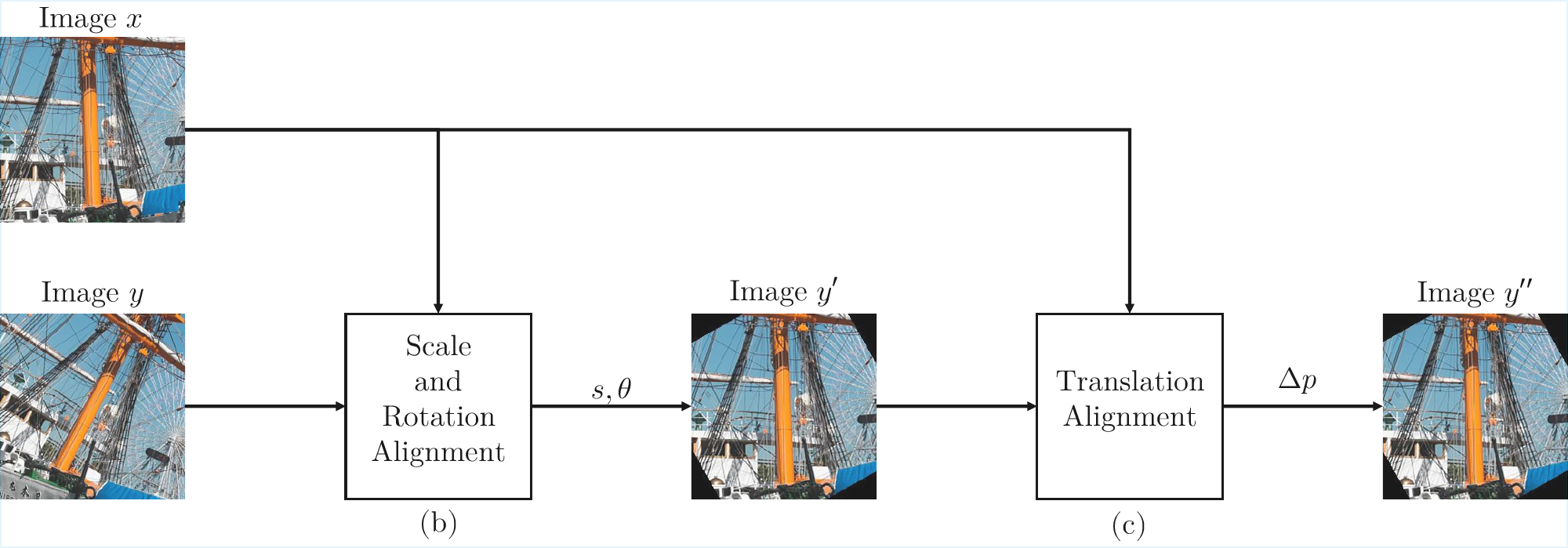}
      \caption{Overview}
      \label{fig:Overview}
  \end{subfigure}
  \begin{subfigure}{\linewidth}
      \centering
      \includegraphics[width=\linewidth]{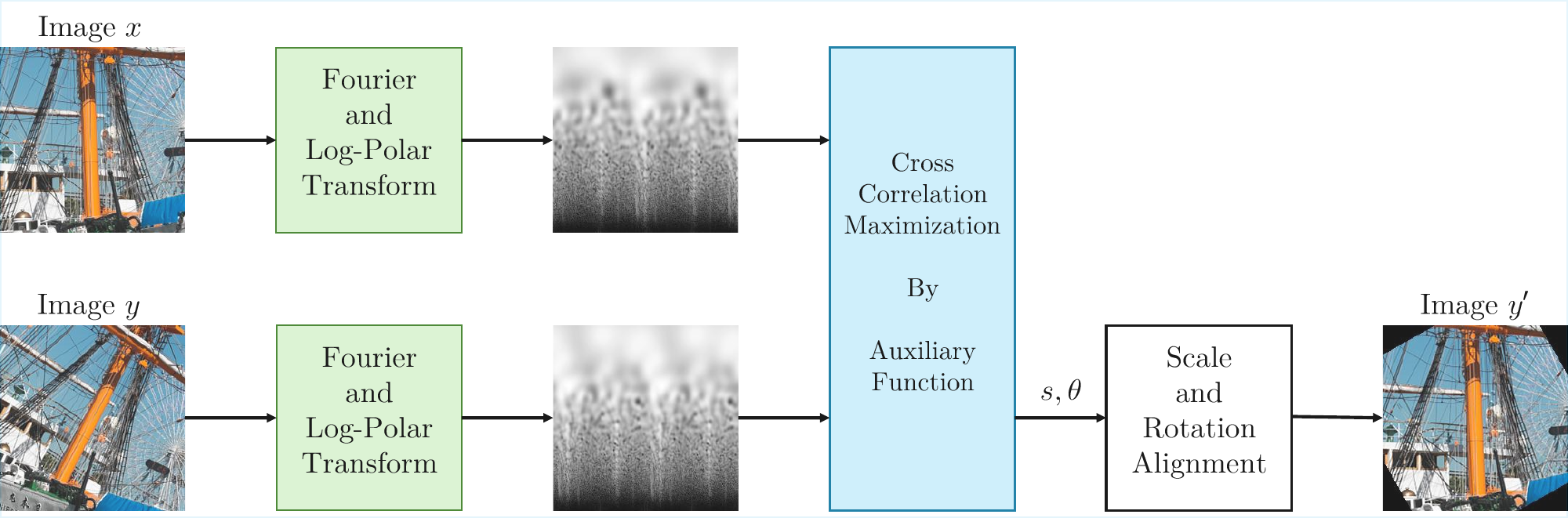}
      \caption{Scale and rotation alignment}
      \label{fig:Scale and Rotation Alignment}
  \end{subfigure}
  \begin{subfigure}[t]{\linewidth}
      \centering
      \includegraphics[width=\linewidth]{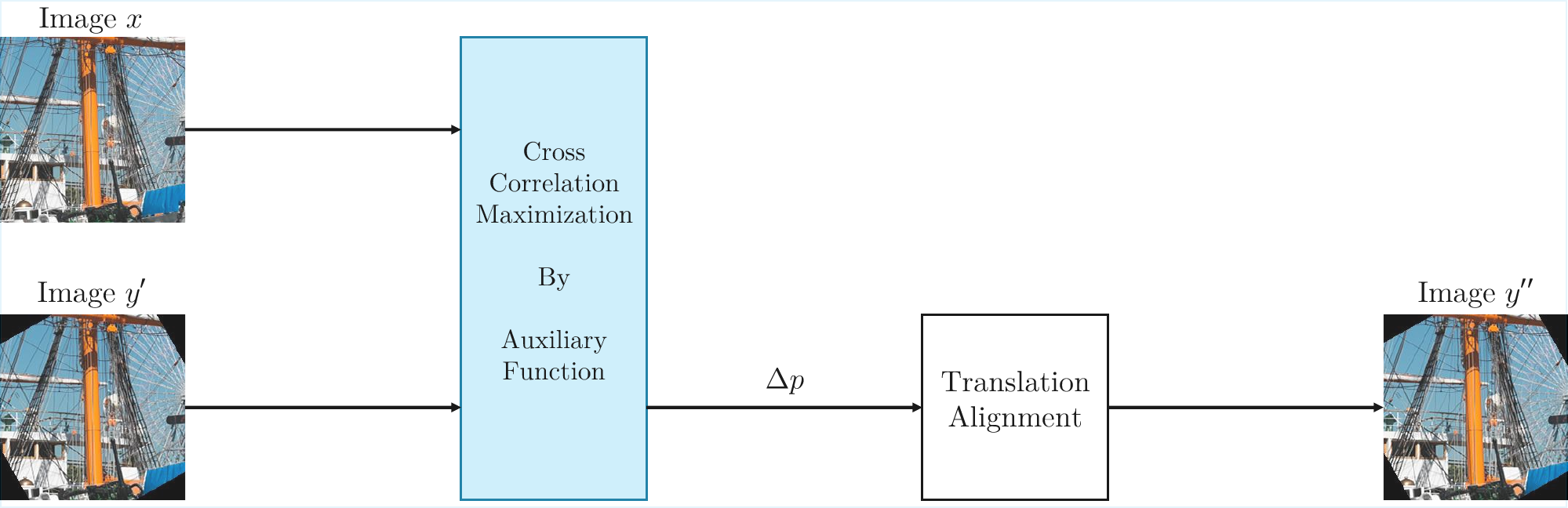}
      \caption{Translation alignment}
      \label{fig:Translation alignment}
  \end{subfigure}
  \caption{Flowchart of proposed method\label{fig:flowchart}}
\end{figure}

The proposed method follows the same two-stage framework.
However, it performs both correlation maximizations using the auxiliary-function-based approach.
As a result, it estimates the similarity transformation parameters $s$, $\theta$, and $\Delta\boldsymbol{p}$ with higher precision, as illustrated in~\Cref{fig:flowchart}.

\subsection{Estimation of scale and rotation}
The Fourier transforms $\hat{x}$ and $\hat{y}$ of the reference image $x$ and the similarity-transformed image $y$ satisfy
\begin{equation}
\hat{y}(\boldsymbol{\omega})
= \frac{1}{s^2}
\hat{x}\left(\frac{1}{s}R(\theta)\boldsymbol{\omega}\right)
\exp\left(j2\pi\boldsymbol{\omega}^\top R(-\theta)\boldsymbol{\Delta p}\right).
\end{equation}
Therefore,
\begin{equation}
    |\hat{y}(\boldsymbol{\omega})|
    = \frac{1}{s^2}\left|\hat{x}\left(\frac{1}{s}R(\theta)\boldsymbol{\omega}\right)\right|
\end{equation}
holds, indicating that the magnitude spectrum $|\hat{y}(\boldsymbol{\omega})|$ of the transformed image is proportional to the magnitude spectrum $|\hat{x}(\boldsymbol{\omega})|$ of the reference image subjected to scaling by $1/s$ and rotation by $R(\theta)$.
In addition,
\begin{align}
    \log \left\| \frac{1}{s}R(\theta)\boldsymbol{\omega} \right\|
    &= -\log s + \log \|\boldsymbol{\omega}\| = -\log s + \rho,\\
    \angle \frac{1}{s}R(\theta)\boldsymbol{\omega}
    &= \angle \boldsymbol{\omega} + \theta = \phi + \theta .
\end{align}
Hence, spatial-domain scaling and rotation correspond to translations in the log-polar representation of the magnitude spectrum along the $\rho$-axis and the $\phi$-axis, respectively.
Accordingly, $s$ and $\theta$ can be estimated by maximizing the cross-correlation function between the magnitude spectra represented in log-polar coordinates $\boldsymbol{r}=(\rho,\phi)^\top$.

The energy of the magnitude spectrum of natural images is often concentrated in low-frequency components. Therefore, directly using the raw magnitude spectra tends to make the correlation dominated by the DC and near-DC components. This reduces the contribution of mid- and high-frequency structures, which are often useful for estimating scale and rotation. To alleviate this effect, the proposed method uses the log-magnitude spectra instead of the raw magnitude spectra. The logarithmic transform compresses the dynamic range of the spectra and reduces the dominance of low-frequency components, leading to more stable estimation in the log-polar domain.
The proposed method estimates $s$ and $\theta$ by maximizing phase-only correlation of the log-magnitude spectra.
Define the log-magnitude spectrum of $x$ on a discrete log-polar grid $\boldsymbol{r}=(\rho,\phi)^\top$ as

\begin{equation}
S_x[\boldsymbol{r}] = \log{\left(|\hat{x}(\boldsymbol{\omega}_{\boldsymbol{r}})| + 1 \right)}, \qquad
\boldsymbol{\omega}_{\boldsymbol{r}} =
\begin{pmatrix}
e^\rho \cos\phi \\
e^\rho \sin\phi
\end{pmatrix}
\label{eq:log_polar}
\end{equation}
where $\rho \in \left\{\log(\min(N,M))\frac{i}{N} \mid i=1,\dots,N \right\}$ and
$\phi \in \left\{\frac{2\pi}{M}i \mid i=0,\dots,M-1 \right\}$.
Because $|\hat{x}[\boldsymbol{\omega}]|$ is defined on a discrete Cartesian frequency grid, bicubic interpolation is used for resampling.
The cross-correlation between $S_x[\boldsymbol{r}]$ and $S_y[\boldsymbol{r}]$ is maximized using the method in Section~\ref{sec:aux}.
In computing the cross-correlation, we subtract the mean of each spectrum to enforce zero mean, and set the weights $w_{kl}=|\hat{\Phi}_2^{(xy)}(\boldsymbol{\omega}_{kl})|^{-1}$ in~\Cref{eq:gcc2_prep}. 

From the estimated translation in the log-polar domain
$\tilde{\Delta r}=(\tilde{\rho},\tilde{\phi})^\top$,
the scale factor $\tilde{s}$ and the rotation angle $\tilde{\theta}$ are computed as
\begin{equation}
\tilde{s}
=
\exp\left(
\log(\min(N,M))\frac{\tilde{\rho}}{N}
\right),
\qquad
\tilde{\theta}
=
\frac{2\pi}{M}\tilde{\phi}.
\end{equation}
Using $\tilde{s}$ and $\tilde{\theta}$, we obtain the scale-and-rotation-compensated image $y'$ as
\begin{equation}
y'[\boldsymbol{p}]
=
y\!\left[\frac{1}{\tilde{s}}R(-\tilde{\theta})\boldsymbol{p}\right].
\end{equation}

\subsection{Estimation of translation}
Using the reference image $x$ and the scale-and-rotation-compensated image $y'$, we estimate the translation $\tilde{\Delta \boldsymbol{p}}$ between $x$ and $y'$.
This is done by maximizing the phase-only correlation between $x$ and $y'$ using the method in Section~\ref{sec:aux}.
As in the previous subsection, we subtract the mean of each image to enforce zero mean and set $w_{kl}=|\hat{\Phi}_2^{(xy)}(\boldsymbol{\omega}_{kl})|^{-1}$ in~\Cref{eq:gcc2_prep}.

Finally, using the estimated translation $\tilde{\Delta \boldsymbol{p}}$, we compensate for the translation in $y'$ to obtain the aligned image $y''$:
\begin{equation}
y''[\boldsymbol{p}]
=
y'[\boldsymbol{p} - \tilde{\Delta \boldsymbol{p}}].
\end{equation}

\section{experiment}
To evaluate the estimation accuracy of similarity transformation parameters achieved by the proposed method, we conducted a simulation experiment.

\subsection{Experimental setup}
\begin{figure}[t]
  \centering
  \begin{subfigure}[t]{\linewidth}
    \centering
    \includegraphics[width=\linewidth]{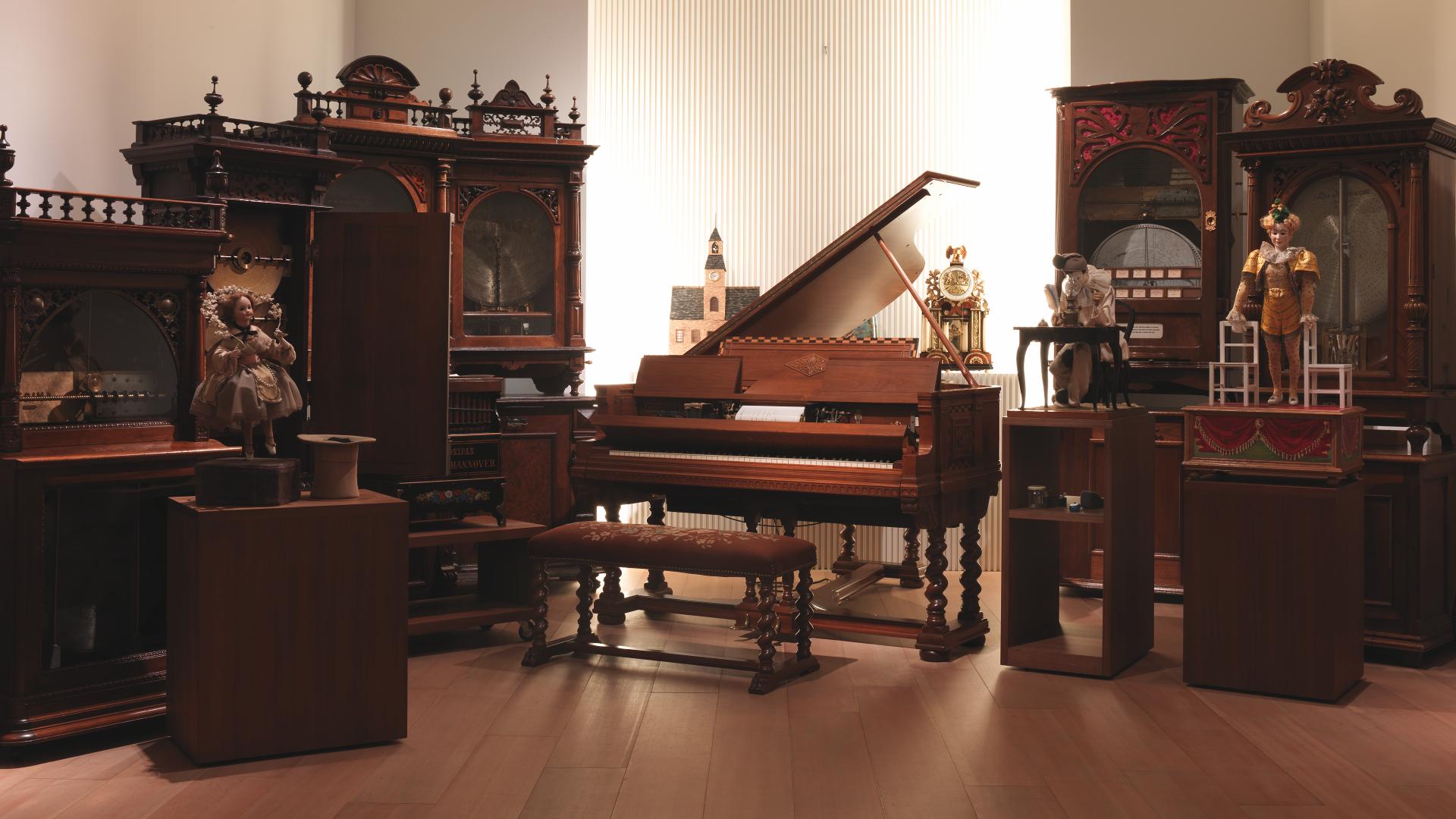}
    \caption{MusicBox}
    \label{fig:MusicBox_original}
  \end{subfigure}
  \hfill
  \begin{subfigure}[t]{\linewidth}
    \centering
    \includegraphics[width=\linewidth]{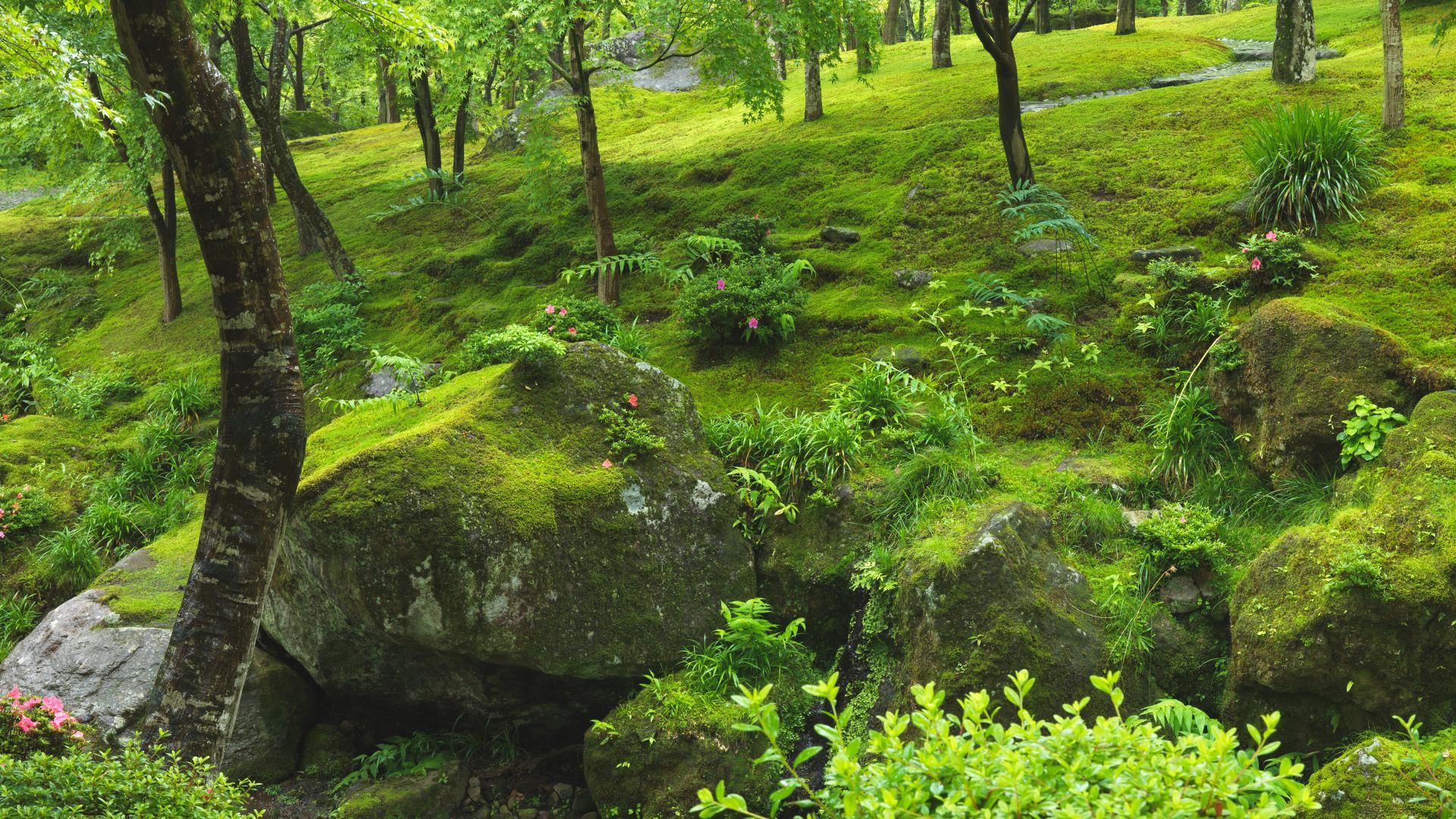}
    \caption{Moss}
    \label{fig:Moss_original}
  \end{subfigure}
  \hfill
  \begin{subfigure}[t]{\linewidth}
    \centering
    \includegraphics[width=\linewidth]{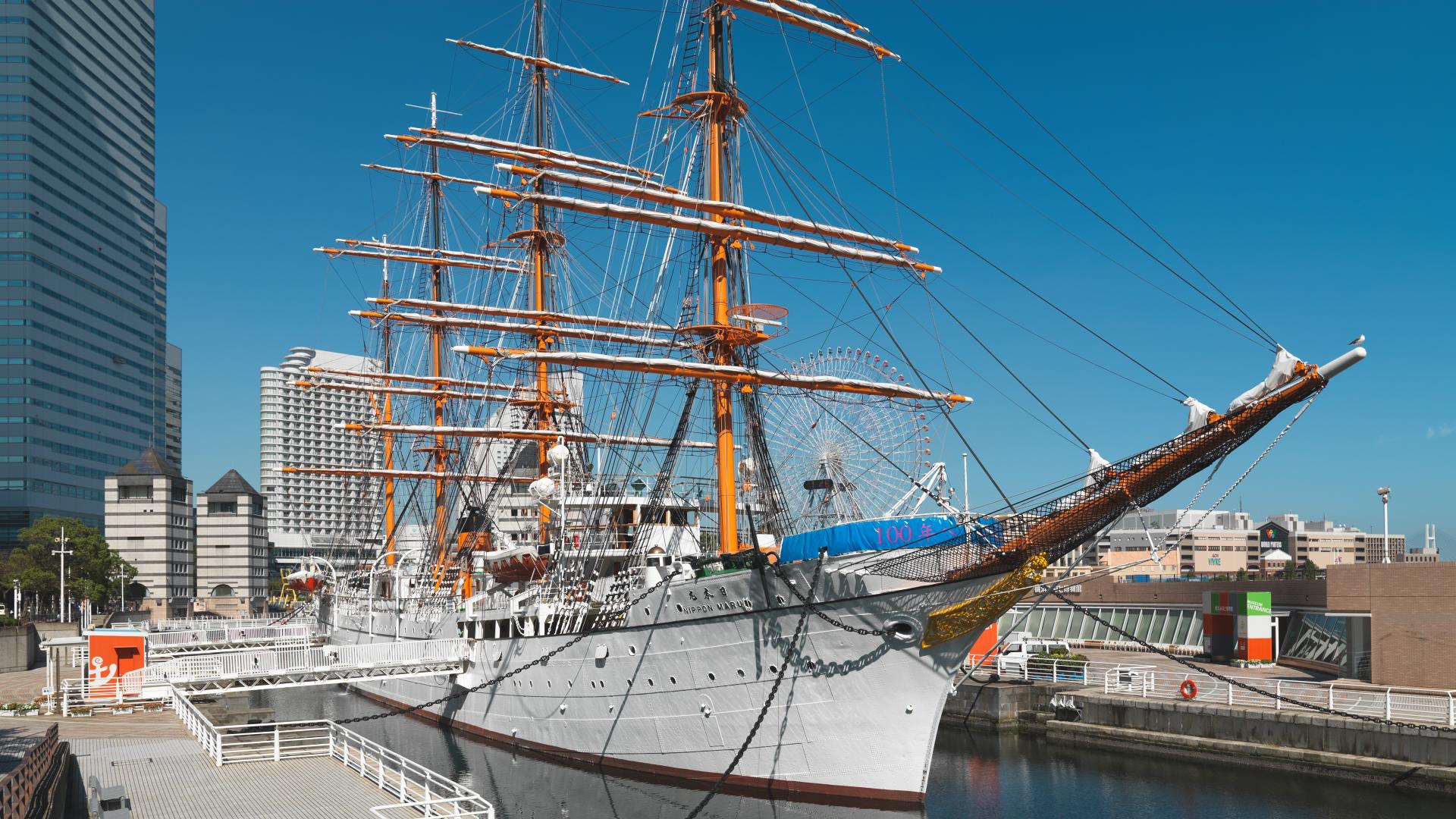}
    \caption{Ship}
    \label{fig:Ship}
  \end{subfigure}
  \caption{Original images used in the experiment\label{fig:original_images}}
\end{figure}

In this experiment, we selected three 2K-resolution images, ``MusicBox'', ``Moss'', and ``Ship'' (see~\Cref{fig:original_images}),
from the ultra-high-definition wide-color-gamut standard image set\footnote{\url{https://www.ite.or.jp/content/test-materials/uhdtv/}}.
For each image, we generated five image pairs by applying a random similarity transformation and then cropping a $256\times256$ pixel region.

\begin{table}[t]
\centering
\caption{Ranges of transformation parameters used in the experiment}
\label{tab:parameter}
\begin{tabular}{c c}
\hline
Parameter & Range \\
\hline
Rotation angle $\theta$ (degrees) & $-30 \sim 30$ \\
Scale factor $s$ & $0.8 \sim 1.2$ \\
Horizontal translation $p_1$ (px) & $-5 \sim 5$ \\
Vertical translation $p_2$ (px) & $-5 \sim 5$ \\
\hline
\end{tabular}
\end{table}

The scale factor $s$, rotation angle $\theta$, and translation $\Delta\boldsymbol{p}$ were sampled from uniform distributions over the ranges listed in Table~\ref{tab:parameter}.
From the generated image pairs, we estimated $s$, $\theta$, and $\Delta\boldsymbol{p}$ using the proposed method.
In computing the DFTs in the proposed method, we applied a Gaussian window whose standard deviation was set to $1/5$ of the input signal length.

As a baseline for comparison, we used a standard Fourier--Mellin-transform-based registration method that estimates translation, rotation, and scale by maximizing discrete cross-correlation.

\subsection{Experimental results}
\Cref{tab:results} lists the absolute errors between the estimated parameters and the ground-truth parameters for the five image pairs generated from each image in~\Cref{fig:original_images}, for both the baseline and the proposed method. The table reports the errors in horizontal translation, vertical translation, scale, and rotation for each image pair, together with their averages and standard deviations. 
\begin{table}[t]
  \centering
  \caption{Absolute errors between the estimated parameters and the ground-truth parameters (Baseline / Proposed)}
  \label{tab:results}

  \begin{subtable}{\linewidth}
    \centering
    \caption{MusicBox}
    \label{tab:Musicbox}
    \begin{tabular}{c c c c c c c c c}
      \hline
      Image & 
      Horizontal shift & Vertical shift & Scale & Angle \\
      \hline
      Image 1 & 0.462 / \textbf{0.005} & 0.121 / \textbf{0.003} & 0.001 / \textbf{0.000} & 0.730 / \textbf{0.023} \\
      Image 2 & 0.080 / \textbf{0.046} & 0.171 / \textbf{0.045} & 0.002 / \textbf{0.000} & 0.581 / \textbf{0.220} \\
      Image 3 & 0.326 / \textbf{0.123} & 0.222 / \textbf{0.110} & 0.007 / \textbf{0.001} & 0.679 / \textbf{0.203} \\
      Image 4 & 0.792 / \textbf{0.083} & 0.073 / \textbf{0.055} & 0.015 / \textbf{0.000} & 0.024 / \textbf{0.020} \\
      Image 5 & 0.037 / \textbf{0.025} & 0.151 / \textbf{0.034} & 0.001 / \textbf{0.001} & 0.532 / \textbf{0.038} \\
      \hline
      Average & 0.340 / \textbf{0.056} & 0.148 / \textbf{0.049} & 0.005 / \textbf{0.001} & 0.509 / \textbf{0.101} \\
      Std. & 0.275 / \textbf{0.042} & 0.050 / \textbf{0.035} & 0.006 / \textbf{0.000} & 0.252 / \textbf{0.091} \\
      \hline
    \end{tabular}
  \end{subtable}

  \medskip

  \begin{subtable}{\linewidth}
    \centering
    \caption{Moss}
    \label{tab:Moss}
    \begin{tabular}{c c c c c c c c c}
      \hline
      Image & 
      Horizontal shift & Vertical shift & Scale & Angle \\
      \hline
      Image 1 & 0.074 / \textbf{0.039} & 0.197 / \textbf{0.019} & 0.006 / \textbf{0.002} & 0.109 / \textbf{0.077} \\
      Image 2 & 0.057 / \textbf{0.016} & 0.030 / \textbf{0.028} & 0.003 / \textbf{0.001} & 0.400 / \textbf{0.234} \\
      Image 3 & 0.157 / \textbf{0.087} & 0.196 / \textbf{0.049} & 0.008 / \textbf{0.001} & 0.255 / \textbf{0.163} \\
      Image 4 & 0.169 / \textbf{0.018} & 0.121 / \textbf{0.008} & 0.001 / \textbf{0.000} & 0.139 / \textbf{0.009} \\
      Image 5 & 0.411 / \textbf{0.054} & 0.291 / \textbf{0.029} & \textbf{0.000} / \textbf{0.000} & 0.263 / \textbf{0.080} \\ \hline
      Average & 0.174 / \textbf{0.043} & 0.167 / \textbf{0.027} & 0.004 / \textbf{0.001} & 0.233 / \textbf{0.113} \\
      Std. & 0.127 / \textbf{0.026} & 0.087 / \textbf{0.014} & 0.003 / \textbf{0.001} & 0.103 / \textbf{0.078} \\
      \hline
    \end{tabular}
  \end{subtable}

  \medskip

  \begin{subtable}{\linewidth}
    \centering
    \caption{Ship}
    \label{tab:Ship}
    \begin{tabular}{c c c c c c c c c}
      \hline
      Image & 
      Horizontal shift & Vertical shift & Scale & Angle \\
      \hline
      Image 1 & 0.426 / \textbf{0.057} & 0.148 / \textbf{0.030} & 0.003 / \textbf{0.001} & 0.368 / \textbf{0.128} \\
      Image 2 & 0.116 / \textbf{0.010} & 0.357 / \textbf{0.026} & \textbf{0.001} / \textbf{0.001} & 0.355 / \textbf{0.188} \\
      Image 3 & 0.496 / \textbf{0.100} & \textbf{0.048} / 0.061 & 0.009 / \textbf{0.002} & 0.013 / \textbf{0.015} \\
      Image 4 & 0.195 / \textbf{0.002} & 0.324 / \textbf{0.036} & 0.006 / \textbf{0.001} & 0.377 / \textbf{0.105} \\
      Image 5 & 0.088 / \textbf{0.015} & 0.160 / \textbf{0.048} & 0.003 / \textbf{0.001} & 0.204 / \textbf{0.028} \\
      \hline
      Average & 0.264 / \textbf{0.037} & 0.207 / \textbf{0.040} & 0.004 / \textbf{0.001} & 0.263 / \textbf{0.093} \\
      Std. & 0.166 / \textbf{0.037} & 0.116 / \textbf{0.013} & 0.003 / \textbf{0.001} & 0.140 / \textbf{0.065} \\
      \hline
    \end{tabular}
  \end{subtable}

\end{table}

From \Cref{tab:results}, the proposed method achieves smaller average absolute errors than the baseline method for all evaluated parameters across all image sets. In particular, the proposed method consistently reduces the average errors in horizontal and vertical translation, scale, and rotation for ``MusicBox,'' ``Moss,'' and ``Ship.'' Although a few individual cases show comparable or slightly larger errors than the baseline, the overall results indicate that incorporating the auxiliary-function-based subpixel maximization into the Fourier--Mellin framework improves the estimation accuracy of similarity transformation parameters.

\begin{figure}[!t]
    \centering
    \begin{subfigure}[t]{0.45\hsize}
      \centering
      \includegraphics[width=\linewidth]{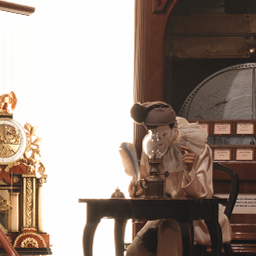}
      \caption{Input image $x$}
      \label{MusicBox: reference image}
    \end{subfigure}
    \begin{subfigure}[t]{0.45\hsize}
      \centering
      \includegraphics[width=\linewidth]{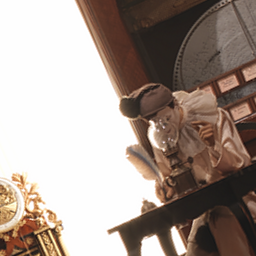}
      \caption{Input image $y$}
      \label{MusicBox: before correct}
    \end{subfigure}\\ \vspace{1em}
    \begin{subfigure}[t]{0.45\hsize}
      \centering
      \includegraphics[width=\linewidth]{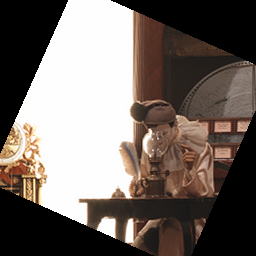}
      \caption{Baseline}
      \label{MusicBox: dcc}
    \end{subfigure}
    \begin{subfigure}[t]{0.45\hsize}
      \centering
      \includegraphics[width=\linewidth]{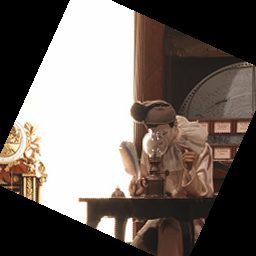}
      \caption{Ours}
      \label{MusicBox: ours}
    \end{subfigure}
    \caption{Alignment result for Image~3, MusicBox\label{fig:MusicBox}}
\end{figure}

\begin{figure}[!t]
    \centering
    \begin{subfigure}[t]{0.45\hsize}
      \centering
      \includegraphics[width=\linewidth]{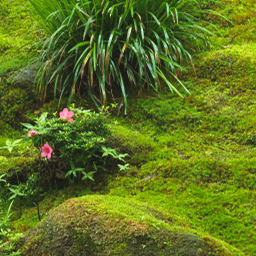}
      \caption{Input image $x$}
      \label{Moss: reference image}
    \end{subfigure}
    \begin{subfigure}[t]{0.45\hsize}
      \centering
      \includegraphics[width=\linewidth]{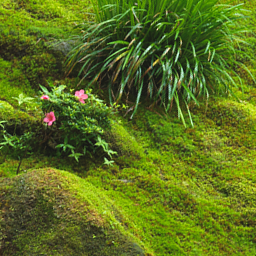}
      \caption{Input image $y$}
      \label{Moss: before correct}
    \end{subfigure}\\ \vspace{1em}
    \begin{subfigure}[t]{0.45\hsize}
      \centering
      \includegraphics[width=\linewidth]{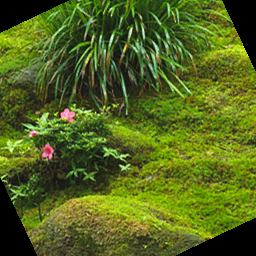}
      \caption{Baseline}
      \label{Moss: dcc}
    \end{subfigure}
    \begin{subfigure}[t]{0.45\hsize}
      \centering
      \includegraphics[width=\linewidth]{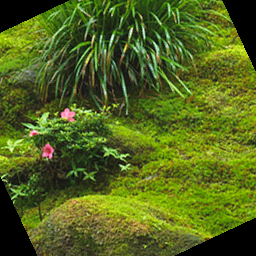}
      \caption{Ours}
      \label{Moss: ours}
    \end{subfigure}
    \caption{Alignment result for Image~4, Moss\label{fig:Moss}}
\end{figure}

\Cref{fig:MusicBox} and \Cref{fig:Moss} show representative alignment examples for Image~3 of ``MusicBox'' and Image~4 of ``Moss,'' respectively. Each figure presents the reference image $x$, the input image $y$, the aligned result obtained by the baseline method, and the aligned result obtained by the proposed method. These examples provide a visual comparison of the alignment quality achieved by the two methods.


This result confirms the effectiveness of maximizing the cross-correlation function with subpixel accuracy using the auxiliary function method within the Fourier--Mellin-transform-based registration framework.

To evaluate the computational cost of the proposed method, we measured the execution times of baseline and the proposed method.
For each method, the total execution time was calculated as the sum of the correlation-peak estimation times in the scale-and-rotation estimation stage and the translation estimation stage.
The machine specifications used in the experiment are listed in Table~\ref{tab:machine_spec}, and the execution time was measured using Python's \texttt{time.perf\_counter()} function.
All computations were performed on the CPU without GPU acceleration.
\begin{table}[tb]
    \centering
    \caption{Machine spec used in the simulation.}
    \label{tab:machine_spec}
    \begin{tabular}{ll}
        \hline 
        Processor & Intel Core i7-1185G7 3.00 GHz \\
        Memory    & 16 GB \\
        OS        & Windows 11 + WSL2 (Ubuntu 24.04 LTS) \\
        Software  & Python 3.12.3\\
        \hline
    \end{tabular}
\end{table}

Table~\ref{tab:total_execution_time} compares the total execution times of baseline and the proposed method.
The reported values represent the mean and standard deviation obtained from the five image pairs generated from each original image.
The proposed method required approximately $ 2.75 $--$ 3.08 $ times the execution time of baseline.
This increase is mainly caused by the iterative auxiliary-function updates, indicating a trade-off between estimation accuracy and computational cost.
\begin{table}[t]
    \centering
    \caption{Total execution time and its standard deviation for correlation-peak estimation (s).}
    \label{tab:total_execution_time}
    \begin{tabular}{lccc}
        \hline
        Method & MusicBox & Moss & Ship \\
        \hline
        Baseline
        & $ 7.9707 \pm 2.7832 $
        & $ 6.6331 \pm 1.3889 $
        & $ 5.7360 \pm 0.9353 $ \\
        Proposed
        & $ 24.512 \pm 7.8969 $
        & $ 18.260 \pm 1.3350 $
        & $ 16.872 \pm 1.4900 $ \\
        \hline
    \end{tabular}
\end{table}

\section{Conclusion}
This paper proposed a method that applies an auxiliary-function-based high-precision correlation maximization algorithm to both (i) scale-and-rotation estimation based on correlation between magnitude spectra in the log-polar domain and (ii) translation estimation based on correlation between images in the spatial domain within a Fourier--Mellin-transform-based registration framework.
This enables high-precision estimation of scale, rotation, and translation, including subpixel-accurate translation estimation.

A simulation experiment confirmed the effectiveness of maximizing the cross-correlation function with subpixel accuracy using the auxiliary function method in Fourier--Mellin-transform-based registration.
Compared with a baseline method that maximizes discrete phase-only cross-correlation, the proposed method achieved smaller estimation errors for scale, rotation, and translation.

Future work includes investigating alternative interpolation schemes for the log-polar resampling step. More extensive comparisons with recent image registration methods, including evaluations of robustness and computational cost, will also be conducted.

\section*{Acknowledgment}
This work was supported by an annual grant from the Tokai University Research and Information Center (TRIC) for the fiscal year 2026.

\printbibliography

\end{document}